# IT ENABLING FACTORS IN A NEW INDUSTRY DESIGN: OPEN BANKING AND DIGITAL ECONOMY

Carlos Alberto Durigan Junior - UNIVERSITY OF SAO PAULO USP - Orcid: https://orcid.org/0000-0003-2185-493X

Kumiko Oshio Kissimoto - UNIFESP - UNIVERSIDADE FEDERAL DE SÃO PAULO/EPPEN - Orcid: https://orcid.org/0000-0002-0316-1307

Fernando José Barbin Laurindo - UNIVERSITY OF SAO PAULO USP - Orcid: https://orcid.org/0000-0002-5924-3782

this paper expects to point out through literature review some IT enabling factors that allow the conception of a new industry design (or governance) specifically in the financial industry illustrated by the cases of the Open Banking and Digital Economy Encourages literature on the field. This paper is structured mostly on literature review, accompanied by results, discussions, and finally, conclusions are presented. It was found five potential enabling factors. The fourth industrial revolution promotes the integration of Information Technology (IT) and strategic resources. New IT demands and uses have been leading to changes in business processes and corporate governance. Better understanding of the IT enabling factors in the cashless economy.

# IT ENABLING FACTORS IN A NEW INDUSTRY DESIGN: OPEN BANKING AND DIGITAL ECONOMY

**Abstract:** *The fourth industrial revolution promotes the integration of Information Technology (IT) and strategic resources. New IT demands and uses have been leading to changes in business processes and corporate governance. Lately, the financial industry has adopted a new integrated banking model known as Open Banking (OB) and the advent of cryptocurrencies has led to the Digital Economy (DE) materialization. Considering these facts, this paper expects to point out through literature review some IT enabling factors that allow the conception of a new industry design (or governance) specifically in the financial industry illustrated by the cases of the Open Banking and Digital Economy. This paper is structured mostly on literature review, accompanied by results, discussions, and finally, conclusions are presented. It was found five potential enabling factors.*

## 1. Introduction, Context, and Problem

In recent years, the world has seen many industries changing their processes and redesigning their governance models due to new Information Technology (IT) solutions or IT assets combinations. Specifically, in the financial industry, it is possible to illustrate the cases of Open Banking and the digital (cashless) economy advent. These observations are inserted in the context of the fourth industrial revolution, which is mostly guided by IT assets and functionalities. Considering these aspects, it is important to identify possible IT enabling factors that may be potential candidates as drivers to allowing a new industrial design and governance.

Information Technology (IT) is important as a driver of product, service, and business innovation in financial markets. There are complex interactions among technology, financial markets, and their stakeholders so the ecosystem's evolution is affected. These forces act as accelerators or decelerators of changes when new IT-enabled innovations have the potential to transform the nature of economic exchange (Kauffman *et al.*, 2015).

According to Laurindo (2008) Information Technology (IT) is a widely accepted term that includes in its meaning; equipment (such as computers, servers, network, communication technology, automation, and network devices), applications, services, human, administrative and organizational aspects (Laurindo, 2008; Porter & Millar, 1985).

IT governance (ITG) is the responsibility of an organization's top management and board of directors; Board-level ITG, defined as "the board's actions to ensure that the IT sustains and extends the organization's strategies and objectives" (Turel & Bart, 2014, p. 224), is an important practice that can positively influence organizational performance (Jewer & McKay, 2012; Nolan & McFarlan, 2005), regardless of the IT use mode of organizations, organization size, sales, and profit orientation of the organization (Turel & Bart, 2014). *et al*.

The fourth industrial revolution allows the combination of numerous physical and digital technologies such as Artificial Intelligence (AI), cloud computing, robotics, augmented reality, additive manufacturing, and the Internet of Things (IoT). Big technological players have entered the digital platform economy. This fosters the ever-changing global economy, based on IT resources and applications. Platforms and the cloud, an essential part of what has been called the "third globalization," reconfigure globalization itself. These digital platforms may disrupt the existing organization of economic activity by changing the logic of value (Kenney and Zysman, 2016).

Technological developments in big data infrastructure, analytics, and services allow firms to transform themselves into data-driven organizations. Due to the potential of big data becoming a game-changer, every firm needs to build capabilities to leverage big data to stay competitive (Lee, 2017).

The advent of digital currencies (cryptocurrencies) has pioneered a new approach to financial transactions; Bitcoin is one of the most known among cryptocurrencies. Blockchain is a Distributed Ledger Technology (DLT) which records transactions made using Bitcoin, it is a type of database that takes several records and puts them in a block, each one is then chained to the next, a cryptographic signature is used in this process. Blockchain works like a ledger, identical copies of a digital ledger are shared with every user (Tasca, 2016)

Moreover, Artificial Intelligence (AI) has an imposing function in this reality; AI mechanisms should consider a variety of topics to establish a transparent and trustful governance model. Due to this new pattern of complex networks imposed by cryptocurrencies, they stand for disruptive changes in the traditional financial governance models. However, their financial impacts over traditional business and assets still have to be more explored and comprehended by literature (Lauterbach and Bonime-blanc, 2016). Table1 below points out some definitions of the terms related to the research of this paper.

Table 1

**Terms and Literature Definition**

| Term | Interpretation |
|---|---|
| Open Banking | According to the Euro Banking Association (EBA, p. 7, 2016) , the term Open Banking can be defined as: "Secure, agile and convenient sharing of products, services, and data from financial sector entities, at the discretion of their clients, for means of opening and integrating IT [Information Technology] platforms and infrastructures of financial service providers". |
| Blockchain | Blockchain is a technology that allows a growing list of data structures (blocks) connected and secured by cryptography (Haber and Stornetta, 1990). In the Blockchain, the distribution of information is decentralized, therefore Blockchain has been a technology able to provide decentralization, immutability, and transparency. Bitcoin, a digital currency (Satoshi Nakamoto, 2008) is the first successful attempt to apply the technology. |
| Digital Economy | According to the European Comission (2013), Digital Economy can be defined as an economy based on digital assets (sometimes named the internet economy) (European Commission 2013). Tapscott (1996) states that Digital Economy is the "Age of Networked Intelligence" where it is "not only about the networking of technology… smart machines… but about the networking of humans through technology" that "combine intelligence, knowledge, and creativity for breakthroughs in the creation of wealth and social development" (Tapscott, 1996) |
| Industry 4.0 | The Fourth Industrial Revolution is related to automated and intelligent production, capable of communicating autonomously with the main corporate players. Industry 4.0 is based on the horizontal and vertical integration of production systems driven by real-time data interchange and flexible manufacturing allowing customized production (Li *et al*., 2017; Thoben *et al*., 2017). |
| Technology Disruption | According to Sood and Tellis (2011) technology disruption occurs when a new technology exceeds the performance of the dominant technology on the primary dimension of performance.<br><br>Danneels (2004) claims that disruptive technology is a technology that changes the bases of competition by changing the performance metrics along which firms compete. |
| IT Enabling Factors | All IT assets, tools, applications, systems, and coordination of IT resources (among others) may allow a new IT governance (industry design). |

Source: Authors

Here the term "*new industry design*" may be understood as new industry governance. This paper intends to identify through literature review some possible IT enabling factors that may permit a new industrial design, specifically in the financial industry illustrated by the cases of Open Banking (OB) and the Digital Economy (DE). This new industry design may be understood as a new governance model of the industry overall.

Considering that the financial industry has seen lately significant processes changes due to IT resources (even in a first step a new IT governance) which leads to a new industry design (or disruption) the main objectives of this paper is to identify through literature review some possible (candidates) IT enabling factors for a new industrial design, specifically in the financial industry applied to the cases of Open Banking (OB) and Digital Economy (DE).

In the literature review of this paper there are eight main topics which are described, as follows; 1-Banks and IT relationship, 2-Open Banking, 3-Blockchain, 4-Cryptocurrencies, 5-Artificial Intelligence, 6-Internet and New Economics, 7-Digital Transformation and 8-IT Governance. It is important to bring some literature references about these topics once they have a relationship with the changes that have been driving the financial industry. Although these eight topics are mentioned, this paper seeks to point out some possible IT enabling factors in a new industry design in the cases of Open Banking and Digital Economy. Following this introduction, this paper is structured in the points; Objectives, Methodology, Literature Review, Findings (along with Discussion, limitations, and future works), and Conclusions.

## 2. Literature Review

The research described in this paper is based on three key-words: Digital Economy, Information Technology (IT) and Open Banking (OB). Considering the new IT governance, new IT business models and new technologies in the financial industry, this paper does consider some literature review about the following topics; 1-Banks and IT relationship, 2-Open Banking, 3-Blockchain, 4-Cryptocurrencies, 5-Artificial Intelligence, 6-Internet and New Economics, 7-Digital Transformation and 8-IT Governance. Among these eight topics, some could be seen as new technologies (tools) and others be understood as new business models or patterns. It is applicable to bring some updated literature references here, once this paper considers IT enabling factors in a new financial industry governance. This literature review is not a systematic review (SLR).

Once this paper considers the IT enabling factors in a new industry design, specifically in financial industry, it is helpful for a better understanding to start pointing out some literature review about the relationship between information technology (IT) and banks. Although banks are not the financial industry as a whole, they can illustrate the relationship well. The item 2.1 of this review brings these relationships.

## 2.1 Banks and IT Relationship

It is important to understand the overall perspective and relationship that the financial industry may have with Information and Information Technology. According to Porter and Millar (1985), banks have high information content in their products and high information intensity in their processes, therefore banks are highly dependent on information. Figure 1 below shows the information intensity matrix from Porter and Millar (1985).

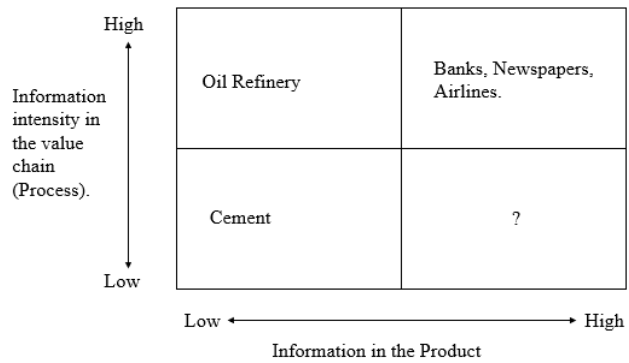

**Figure 1.** Information intensity matrix Porter e Millar (1985).
Source: Adapted from Laurindo (2008) Porter & Millar (1985).

McFarlan (1984) brought to the literature the Strategic Grid, this enables the positioning of industry according to the alignment of both current and future impact that Information Technology applications have on a firm strategy. According to the Grid, it is possible to realize that banks have are very dependent on IT applications, thus IT significantly affects the strategy of banks. Figure 2 below shows the Strategic Grid of McFarlan.

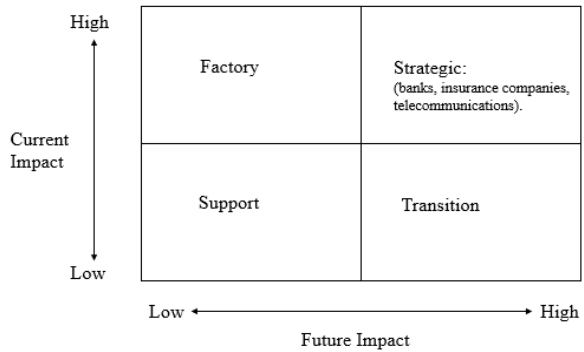

**Figure 2.** Strategic Grid of McFarlan (1984).
Source: Adapted from Laurindo (2008), Fernandes & Kugler (1990), McFarlan (1984).

Henderson and Venkatraman (1993) proposed to the literature a model of the strategic alignment between IT and firm strategy. The model considers both internal and external factors. It contains four macro classifications (illustrated in Figure 3 below). In this model it is possible to realize the direction of a strategy; sometimes IT may define a firm strategy, others this last one can drive the IT strategy (vice-versa). Banks and fintech, as highly influenced by new IT tools, assets, and applications, might be best classified in the Technological Transformation and Competitive Potential squares.

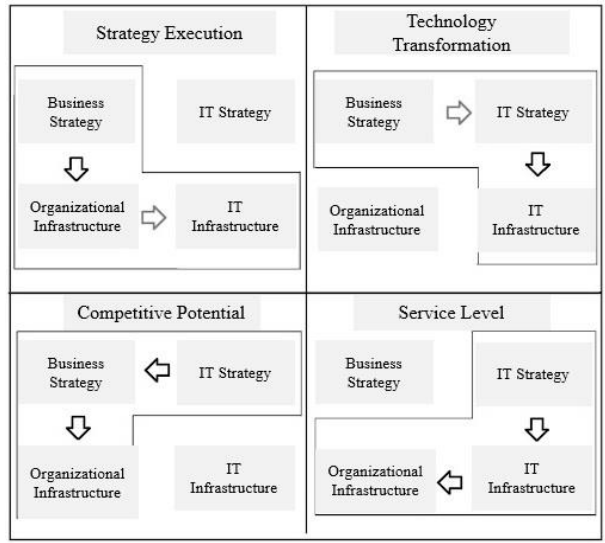

**Figure 3.** Strategic Alignment Perspectives of Henderson & Venkatraman (1993).
Source: Adapted from Laurindo (2008), Henderson & Venkatraman (1993)

**2.2 Open Banking:**

Open Banking represents an innovation with a disruptive potential to integrate banking institutions, fintech, and other players, to unite market-opening initiatives, allowing agility and uniformity to the procedures adopted by banks. (Ribeiro and Bagnoli, 2020).

Open Banking refers to the use of open Application Programming Interfaces (APIs) that allow third-party developers to create applications and services in banks (Ribeiro and Bagnoli, 2020). According to the Euro Banking Association (EBA, 2016, p. 7), the term Open Banking (OB) can be defined as "Secure, agile, and convenient sharing of products, services, and data from financial sector entities, at their discretion of its customers, by opening and integrating IT (Information Technology) platforms and infrastructures of financial service providers".

Open banking can promote the maximized benefits of customers through data sharing and deeper cooperation between financial institutions (Avital *et al*., 2017). Open banking may transform the financial industry by breaking the data silos of traditional banks. Open banking dynamics allow fintech and other players to access customer data. The Consultive Group to Assist the Poor (CGAP) has identified twelve design components that policymakers need to consider when developing a framework for an open banking regime are: 1. Types of services; 2. Participants; 3. Types of data; 4. Payment initiation; 5. Mandatory vs. voluntary; 6. Technical specifications for data sharing; 7. Staged implementation; 8. Lead regulator/policy mandate; 9. Governance; 10. Cost distribution; 11. Data privacy and portability; 12. Liability and consumer protection (CGAP, 2020).

Open Banking is a collaboration model in which bank data is shared via APIs between two or more parties. APIs are Application Programming Interfaces that enable interaction between other applications. Its use in the financial sector has attracted numerous non-banking companies, which are contributing to the creation of the digital financial ecosystem (Freire e Domingos, 2020).

With Open Banking, there are many functions or procedures used by computers to access operating system services, software libraries, or other systems. The APIs describe operations, inputs, and outputs of software components, enabling programmers to understand how to use pieces of software without knowing the internal algorithms, following rules that stipulate inputs and outputs, as well as allowing computer applications to communicate over the network, using a language in which both interact. (Ribeiro and Bagnoli, 2020).

According to Barros (2018) 7% of banks had planned to invest in open APIs and 73% had been willing to open their APIs to third-party developers, according to the *"2018 Global Payments Insight Survey: Retail Banking"* survey by ACI Worldwide and Ovum (Barros, 2018). Traditional banks are welcoming fintech, favoring the occurrence of Seamless banking (bank available anywhere and in any application) (Ribeiro and Bagnoli, 2020).

In the context of the digital platform, open APIs are considered as the boundary resources through which organizations can share a core functionality based on a software platform and provide the opportunity to external developers to produce modules that interoperate with it (Tiwana *et al.*, 2010; Ghazawneh and Henfridsson, 2013). API can help cultivate and expand the ecosystem and invite new actors as well as knowledge and capabilities from outside the firm (Van de Ven, 2005).

## 2.3 Blockchain

Blockchain is a technology that allows a growing list of data structures (blocks) connected and secured by cryptography (Haber and Stornetta, 1990). In the Blockchain, the distribution of information is decentralized, therefore Blockchain has been a technology able to provide decentralization, immutability, and transparency. Bitcoin, a digital currency (Satoshi Nakamoto, 2008) is the first successful attempt to apply the technology.

The World Economic Forum (WEF) published a report in August 2016 named "The future of financial infrastructure an ambitious look at how Blockchain can reshape financial services". Blockchain systems use a Distributed Ledger Technology (DLT) and the WEF focused on topics in the financial markets where this technology can be applied to optimize the process and reduce costs. For instance, trade finance, global payments, and assets clearing/settlement are some evidence where this technology can be considered (WEF, 2016).

Blockchain-based solutions work well for both payments and settlement frameworks, using decentralized protocols. It is possible to make international payments and combine any currency. Transactions can be settled directly between the parties. According to Tasca (2016) Bitcoin seems to have been used more to transfer a large amount of money from person to person rather than used as payment for general consumptions (Tasca, 2016).

Considering governance, both legal code and technical code (software/hardware) may regulate general aspects. The impact of both must be considered in setting out regulations that cover distributed ledger systems. Lehdonvirta and Ali (2016) pointed out distinctions between governance (rulemaking by the owners or participants of a system to safeguard their private interests) and regulation (rulemaking by an outside authority tasked with representing the interests of the public) (Lehdonvirta and Ali, 2016).

Topics related to Information Technology, such as computing and cryptography have led to infrastructures that allow disintermediated and decentralized markets. Tasca (2015) declares that there are at least nine possible mechanisms into which Blockchain technology might be considered, are: 1) intermediation; 2) clearing and settlement; 3) recording systems; 4) rating or voting systems; 5) database systems; 6) distributed storage, authentication, anonymization of private information; 7) rewarding and punishing incentive schemes; 8) transaction traceability schemes; and 9) refereeing, arbitration, or notarization (Tasca, 2015).

## 2.4 Cryptocurrencies

According to Tasca (2015), cryptocurrencies can be defined as: *"Money expressed as a string of bits sent as a message in a network that verifies the authenticity of the message via* *different mechanisms, such as proof-of-work (PoW) or proof-of-stake (PoS)"*. For accountability, every transaction needs to be transparent. Anonymity is preserved, however, all transactions are traceable regarding the fact that they are recorded in a public ledger (Tasca, 2015).

Tasca (2016) points out that economic theory defines money by looking at its functions as a medium of exchange, a unit of account, and a store of value. The author adds a fourth monetary aspect which was named as "*Transactional utility of reward*". This last aspect is related to the current digital era, which is characterized by a cashless and massively connected society utilizing high-frequency transnational transactions of products and services that have been being more digitalized. The transactional utility of reward gains relevance supported by global digital wallets (Tasca, 2016).

Cryptocurrencies are decentralized digital currencies; Bitcoin is the most known one. In more detail, a cryptocurrency is a digital token that exists within a particular system which generally consists of a P2P network, a consensus mechanism, and a public key. A cryptocurrency has three main properties; Digital barrier asset, Integrated payment network, and also non-monetary use cases.

There are many interdependencies in economic systems, they involve from simple local transactions to global large investment networks, single or clustered. The systemic complexities of economic networks depend not only on micro factors but also are influenced by macroeconomic forces. In the literature, there are two principal approaches to study economic networks: socioeconomics and complex systems (Schweitzer et. al, 2009). For instance, Bitcoin is a tiny fraction of the global economy, yet its network can disrupt the existent global economic governance model (Carlson, 2016).

According to the Bank for International Settlements (BIS), if Central Banks use cryptocurrencies to consumers and firms, this could significantly affect core banking areas such as payments, financial stability, and monetary policy (BIS, 2018).

**2.5 Artificial Intelligence (AI)**

Lauterbach and Bonime-blanc (2016) stated the definition of Artificial Intelligence (AI), according to the AI is focused on machine learning and software specification to solve problems. There are two approaches for AI; the first is embedded into specific tasks and the second is devoted to general comprehension (deep AI or AGI). AI can carry out issues related to different fields, such as decision-making processes, visual perception, and speech recognition (Lauterbach and Bonime-Blanc, 2016).

The authors point out the most important points of AI which have supported its critical importance, are: enabling the growth of data, algorithms improvement, cloud technology, smart networks, helping cyber defense, and decreasing coding mistakes (Lauterbach and Bonime-blanc, 2016).

According to Bostrom (2016), the literature about innovation economics is relevant, yet it remains inconclusive about the topic, additionally, the openness in AI may have a positive effect on the rate of technological advances in the field. Regulation and cultural norms may also interfere in the processes of AI and innovation. The openness in governance structure may also foster trust and transparency in values and common goals shared by stakeholders (Bostrom, 2016).

Cloud computing has been used based on the type of availability of cloud computing resources and accessibility. When IT infrastructure offering these services is located at a secured site and both staff & customers accessing it from various remote locations – data privacy and systems security continue to be with no tolerance for risk. With new technologies, interconnection of various devices, mobile devices, social networks, data, and different regulatory norms in various countries makes the security framework for cloud architecture more complex and subject to critical evaluation. Data privacy and Data security are related (Mahalle *et al.,* 2018).

Borges *et al*. (2021) brought to the literature some evidences that there is a connection between Artificial Intelligence (AI) and business strategy. Following the literature review the authors claimed that AI is mainly strategically used in three ways; (i) Related to decision making processes; (ii) Able to improve customers relationship and engagement decisions; and (iii) AI enables communication among machines (Borges *et al*., 2021).

**2.6 Internet and New Economics**

Many other terms and concepts are used to refer to the new economy. According to Gereffi (2001), many refer to this new situation as "digital economy", "innovation economy", "network economy", "electronic economy" (Gereffi, 2001). The digital economy is widely analyzed by Tapscott (2001) with an emphasis on the internet as an enabling agent for business networks that opens new paths for competitive strategy (Tapscott, 2001). For Porter and Millar (1985), the definition of IT must be understood broadly, that is, in addition to encompassing all the information created and used by businesses, it must contemplate the large spectrum of increasingly converging and interconnected technologies (Porter and Millar, 1985).

The new economy, according to Corrêa and Corrêa (2006), is the term used in academic and business circles to reflect the reality of the economy and the dynamics of the market, strongly influenced by globalization and new technologies, whose main products and assets are information and knowledge (Correa and Corrêa, 2006). The role of networks in internalization theory has evolved over the years. digital platforms emphasize the economics of networks and particularly increasing returns to scale (Banalieva and Dhanaraj, 2019). Naisbitt (1982) and Synnott (1987) highlighted the information society as an economic reality that has a profound impact on business and competition (Naisbitt, 1982 and Synnott, 1987).

**2.7 Digital Transformation**

In recent years, digital transformation (DT) has emerged as an important phenomenon in strategic IS research (Bharadwaj *et al*., 2013; Piccinini *et al*., 2015). At a high level, DT encompasses the profound changes taking place in society and industries through the use of digital technologies (Agarwal *et al*., 2010; Majchrzak *et al*., 2016). In line with previous findings on IT-enabled transformation, research has shown that technology itself is only part of the complex puzzle that must be solved for organizations to remain competitive in a digital world (Vial, 2019).

In the latter, DT is depicted as a higher-level phenomenon that disrupts the competitive environment and demands a response from the part of the organization literature that refers to two novel concepts in the context of DT: digital business strategy and digital transformation strategy. Bharadwaj *et al*. (2013) argue that digital technologies call for researchers to study the fusion between organizational strategy and IS strategy (Kahre *et al*., 2017) rather than their alignment Westerman *et al*. (2011) They view DT strategy as separate from "IT strategies and all other organizational and functional strategies" while structural changes, defined as "variations in a firm's organizational setup" (Westerman, *et al*., 2011, p.340-341).

Open Strategy (OS) research explores boundaries in strategy, Information Systems (IS), and the function of technology in strategy work. OS is established when organizations include internal and external stakeholders in strategizing, gaining access to new insights which inform strategic direction and company transformation (Whittington *et al*., 2011; Hautz *et al*., 2017;

Mount *et al*., 2020). When studying OS, it is important to take into consideration the IT-enabled practices and therefore the digital work of top managers in OS (Mortona *et al*., 2020).

About IS Tavakoli *et al*. (2017) contributed to the theoretical development of OS by explicating that IT is essential to complementing or replacing analog strategy work, and in enabling openness occurrence (Mortona *et al*.,2020). This highlights the powerful combination of OS and IT and its capacity to impact strategy development and organizational transformation (Hautz *et al*., 2017).

## 2.8 IT Governance

IT Governance (ITG) is essential to an organization's success (Weill & Ross, 2005). As IT is associated with risk and value opportunities, ITG has become imperative for business organizations to meet the challenges presented by the business environment, IT stands for a competitive advantage. There are many studies about ITG (e.g., Calder & Moir, 2009; Calder, 2005; Willson & Pollard, 2009), and some reports from leading enterprises in ITG such as ITGI and ISACA. However, these studies are not at the same level of ITG importance in some aspects such as Critical Success Factors (CSFs) (Alreemy, *et al*., 2016).

COBIT 5 introduced CSFs for the ITG processes but they cannot be used as CSFs for the entire ITG implementation. ISO 38500 introduced six principles, it was the first international standard explicitly addressed the governance of ICT. The alignment between IT and business is an important aspect that should be considered in the implementation of IT (Alreemy, *et al*., 2016).

The Information Systems (IS) literature related to inter-firm IT governance deals primarily with the IS outsourcing environment Inter-firm IT governance usually includes contractual governance and relational governance (Cao, Mohan, Ramesh, & Sarkar, 2013). Both relational and contractual governance is necessary and effective governance mechanisms in the process of managing IT outsourcing (Deng, Mao, & Wang, 2013; Kim, Lee, Koo, & Nam, 2013; Tiwana, 2010) (Chi *et al*., 2017).

Lazic *et al.* (2011) found that IT governance is positively related to business performance through IT relatedness and business process relatedness. Prasad *et al* (2012) suggest that IT governance structures contribute to firm performance through IT-related capabilities which improve the effectiveness and efficiency of the internal business processes. Yet, among these few works on the governance–performance link, there is no consensus as to exactly how IT governance enhances performance and it is still unclear by which precise mechanisms IT governance exerts its effects on firm performance (Ju Wu *et al*., 2015).

Board-level ITG matters for firms and their performance. To narrow this theoretical gap, it is possible to rely on the RBV, dynamic capability perspective, organizational contingency theory, and reactance theory. Many studies on the value of IT adopt the RBV to explain how organizations gain competitive advantage and superior performance by utilizing IT. RBV defines that organizations hold resources and capabilities, which allow them to achieve competitive advantage and improve organizational performance (Barney, 2001).

Prior research has identified and defined different kinds of IT capabilities (Bharadwaj, 2000; Bharadwaj *et al*., 1999; Bhatt & Grover, 2005; Feeny & Willcocks, 1998; Mithas, Ramasubbu, & Sambamurthy, 2011; Wade & Hulland, 2004), which is one family of organizational capabilities. More specifically, IT capabilities center on IT resources and practices and have been shown to improve organizational performance (Bharadwaj, 2000; Chakravarty, Grewal, & Sambamurthy, 2013, *Apud* Turel *et al*., 2017).

To have a positive effect on an organization's competitive position and performance, IT capability needs to be valuable, heterogeneously distributed, and imperfectly mobile (Mata, Fuerst, & Barney, 1995). Strategic alignment is defined as the degree to which the mission, objectives, and plans contained in the business strategy are shared and supported by the IT strategy (Reich & Benbasat, 1996). Moreover, organizational performance is one outcome of this strategic alignment (Henderson & Venkatraman, 1993; Sabherwal & Chan, 2001; Wu *et al.*, 2015) (Turel *et al.*, 2017).

IT Governance IT Governance is related to Corporate Governance and is concerned with the control and transparency of decisions in Information Technology, without disregarding mechanisms and processes to increase the effectiveness of IT (Peterson, 2004). The components of the Information Technology environments, such as networks, servers, and applications, are increasingly complex, composed of more sophisticated items, and require a higher level of technological integration. The increasing integration of administrative and factory functions allows the creation of Enterprise Resource Planning (ERP) and Manufacturing Execution Systems (Amess *et al.*, 2007).

The integration between suppliers and customers, with the creation of integrated supply chains, and the intensification of the relationship with customers, helps to increase the integration, sophistication, and complexity of the IT environment (Fernandes; Abreu, 2008).In the view of Webb *et al*. (2006), the three elements most frequently associated with IT Governance in the literature are structures, control frameworks, and processes. However, defining IT Governance involves a broader spectrum. The establishment of a "definitive definition" that observes such a "broad spectrum" and that does not limit or restrict the conceptual scope (Webb *et al*., 2006).

## 3. Methodology

Keywords were used to search papers in the field, some academic sources were used (Google Scholar, Web of Sciences, Scopus, Science Direct. More than one hundred papers were found, these seventy papers were selected, analyzed, and grouped into eight themes according to the Literature Review in this paper. (Banks and IT relationships, Open Banking, Blockchain, Cryptocurrencies, Artificial Intelligence, Internet and New Economics, Digital Transformation, and IT Governance). The criterion for choosing the papers is to observe some relationship between the keywords and the financial industry changes. The methodological approach considered in this paper is not a systematic literature review (SLR).

The main goal while reading the articles was to observe IT enabling factors (all IT assets, tools, applications, systems, and coordination of IT resources, among other possibilities) which could allow a new industrial design in the cases of Open Banking and Digital Economy (this is the criteria of inclusion). Here the term "new industry design" may be understood as new industry governance. No statistical tool was employed. Figure 4 below states the methodological steps applied.

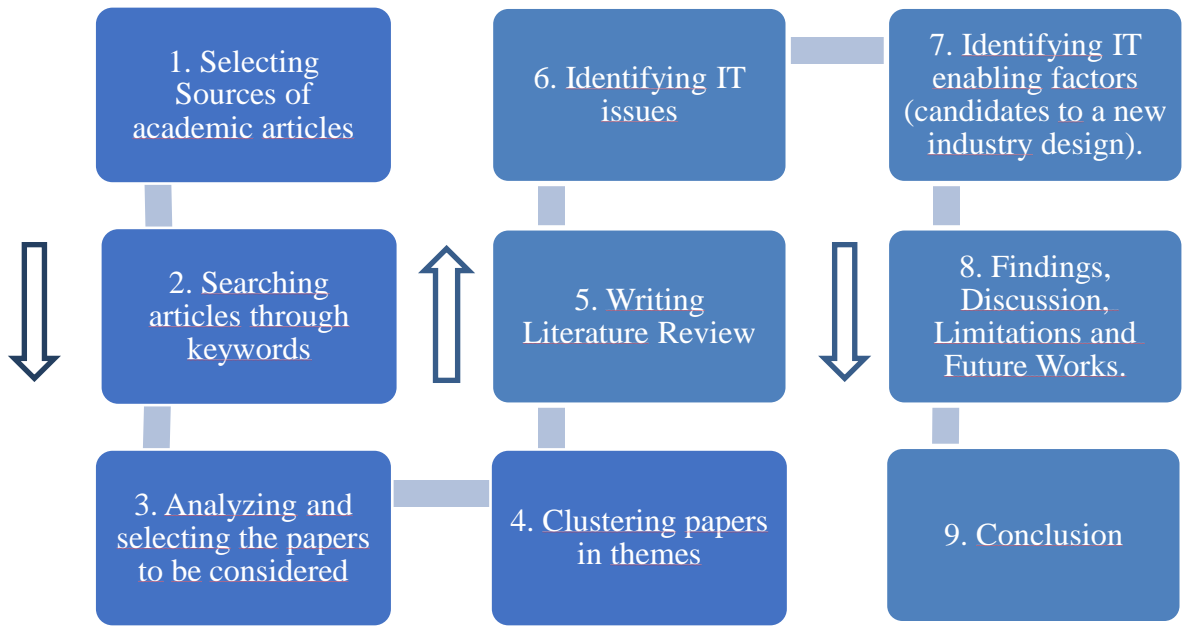

**Figure 4.** Methodological flow.
Source: Authors

Table 2 below shows details of each methodological step.

Table 2

## Methodological details

| Step | Details |
|---|---|
| 1 | It was used Web of Sciences, Scopus, Google Scholar, and Science Direct. |
| 2 | The keywords of this paper are: Digital Economy, Information Technology (IT), IT Strategy, and Open Banking. Other terms that were searched in this paper were: Cryptocurrency, Artificial Intelligence, and IT/ Technological Transformation. |
| 3 | The papers considered were the ones that could help answer the research question of this paper (finding/ identifying IT enabling factors candidates for a new industry design) along with the ones that allowed literature review in the fields. |
| 4 | The papers were clustered into eight themes (Banks and IT classification, Open Banking, Blockchain, Cryptocurrencies, Artificial Intelligence, Internet and New Economics, Digital Transformation, and IT Governance). |
| 5 | A literature review was made for each theme (group or cluster named above) |
| 6 | IT issues were identified to get the next step (number eight) |
| 7 | IT enabling factors were identified and listed. |
| 8 | Discussion is made. |
| 9 | Conclusions are presented. |

Source: Authors.

**4 Findings, Discussion, Limitations, and Future Works**

According to the literature review presented in this article, it is possible to point out some IT potential candidates for enabling factors to a new industrial design in the cases of Open Banking and Digital Economy. In the papers reviewed, were found one main factor for Open Banking and four factors for Digital Economy, they are presented in Tables 3 and 4 below. It is important to mention that they are being presented as potential candidates, in this paper these factors were not empirically tested to prove them as enabling factors in a new industrial design.

Table 3

**Potential IT enabling factor for Open Banking**

| Enabling Factor | Literature | Consequences/ Uses |
|---|---|---|
| API (Application Programming Interface) | Ribeiro and Bagnoli (2020) | <ul><li>Allow third-party developers to create applications and services in banks.</li><li>use pieces of software without knowing the internal algorithms.</li><li>allowing computer applications to communicate over the network.</li></ul> |
| API (Application Programming Interface) | Freire and Domingos (2020) | <ul><li>Collaboration and data sharing among banks.</li><li>Digital Financial Ecosystem.</li></ul> |
| API (Application Programming Interface) | Van de Ven (2005) | <ul><li>Expand the ecosystem and invite new actors.</li><li>Knowledge and capabilities from outside the firm.</li></ul> |
| API (Application Programming Interface) | Tiwana *et al*. (2010) Ghazawneh and Henfridsson (2013) | <ul><li>Share a core functionality between organizations.</li><li>Module's interoperation.</li></ul> |

Source: Authors.

Table 4

**Potential IT enabling factors for Digital Economy**

| Enabling Factor | Literature | Consequences/ Uses |
|---|---|---|
| Blockchain (DLTs) | Haber and Stornetta (1990) | <ul><li>Technology that allows a growing list of data structures.</li></ul> |
| Blockchain (DLTs) | Tasca (2016) | <ul><li>Possibility of making international payments and combine any currency.</li><li>Use of decentralized protocols.</li></ul> |
| Computing & Cryptography | Tasca (2015) | <ul><li>Decentralized Infrastructure.</li><li>Infrastructures that allow disintermediated and decentralized markets.</li></ul> |
| Global Digital Wallets | Tasca (2016) | <ul><li>Transactional Utility</li></ul> |

Source: Authors.

Application Programming Interface (API) is a potential candidate for an IT enabling factor in a new industrial design in the case of Open Banking, according to Ribeiro and Bagnoli (2020); Freire and Domingos (2020); Van de Ven (2005); Tiwana *et al*. (2010) Ghazawneh and Henfridsson (2013). API application using allow functionalities that may lead to new governance, once there is a new IT governance model, the governance of the industry itself might chance as well; thus that is the reason to refer to it as a new industrial design.

The use of API enables financial firms to change data, promote systems interoperability, communication, and collaboration among them. As it is shown in table three above. As Porter (1985) stated, technological and digital innovation has been credited with having significant strategic implications for firms by shifting the competitive landscape and changing the market dynamics in an industry (Porter, 1985). The distinction between internal and external APIs helps us to better understand the benefits of API technology in terms of systems integration and data sharing within and across firms. Scholars suggest that at the heart of systems integration are the principles of interoperability and modularity (Bodle, 2011).

For the Digital Economy it is possible to point out that Blockchain (DLT), Computing, Cryptography, and Digital Wallets are possible candidates for IT enabling factors in a new industrial design. Among them, mainly Blockchain is a Distributed Ledger Technology that significantly changes the governance model, once this technology allows decentralization and disintermediation (Tasca, 2016).

Blockchain can be seen as a continuously growing list of records managed by a peer-to-peer network and often combined with artificial intelligence cloud computing big data Internet of Things (IoT) and other technologies (Wang *et al*. 2020). According to Wang *et al*. (2020) banks need to redesign their system, adapt to collaborative work between blockchains and other huge bank applications. (Wang *et al*., 2020). The observations above do confirm the positioning of banks as highly dependent on IT in the Information intensity matrix Porter e Millar (1985) and the Strategic Grid of McFarlan (1984) (Figures 1 and 2, in the Literature Review).

Considering the strategic alignment perspective based on Henderson and Venkatraman (1993) it would be necessary a detailed analysis to better point out where to classify in the most suitable perspective. However, the use of new IT tools as potential enabling factors in new industry governance would generally fit either on the technological transformation or on competitive potential (Figure 3 in the Literature Review). The literature describes digital technologies as inherently disruptive (Karimi and Walter, 2015). Recently it has occurred the emergence of platform business models that move away from the traditional vertical integration of the firm and introduce a flatter, more inclusive, and innovation-centric approach to value creation (Gawer 2009). It is observable that IT resources to allow new industry governance.

## 5 Conclusions

Firstly, about limitations and Future works, one limiting factor that this study has is the fact that it is not based on a Systematic Literature Review (SLR), so there could be a lack of potential references which bring light on the purpose of this paper, thus this paper may be missing other potential IT enabling factors for new industry governance. Another factor is that this paper neither set criteria to present an IT enabling factor nor empirically test them, that is why they are presented as potential candidates. Considering these limiting factors, future works may consider a SLR literature review and empirically (or statistically) test the potential IT enabling factors.

This paper provides a Literature review about the main important pillars which have some relationship with the changes that the financial industry has seen lately. The main objective was to identify some potential IT enabling factors that allow the new governance that this industry has been adopting. According to the papers reviewed, the literature shows that Application Programming Interface (API) is a potential candidate for being an IT enabling factor in the new governance related to Open Banking. In the case of the Digital Economy, this work shows that Blockchain, Computing, Cryptography, and Global Digital Wallets are potential IT candidates for allowing the new governance.

Although this paper did not use a SLR literature review, future works may consider this last approach and empirically (or statistically) test the potential IT enabling factors. This work contributes to literature once it considers the roots related to IT that allows the new design that the financial industry has adopting. Moreover, this study takes into consideration that IT governance models are related to a complex networked business model (which may share governance models among its players). This study makes one thinks about an IT Governance aspect applied not only to a player but indeed to a complex networked structure with different players. This needs IT enabling factors to promote interaction among many actors.

Future studies should take these points into consideration, in order to improve the literature review and also perform qualitative and/or quantitative researches in organizations.